\documentclass{iau} 
\usepackage{graphicx}
\usepackage{natbib}
\newcommand{\iso}[2]{\hbox{${}^{#1}{\rm #2}$}}
\newcommand{\msun}{\ensuremath{{M}_{\odot}}}

\title[Stellar yields 
and abundances] 
{Stellar yields 
and abundances:\\ new directions from planetary nebulae}

\author[Maria Lugaro, et al.]
{Maria Lugaro$^{1,2}$
 \and Amanda I. Karakas$^{2,3}$
\and Marco Pignatari$^{4,5}$
\and Carolyn L. Doherty$^1$
}

\affiliation{$^1$Konkoly Observatory, Research Centre for Astronomy and Earth 
Sciences,\\
Hungarian Academy of Sciences, H-1121 Budapest, Hungary \\email: 
{\tt maria.lugaro@csfk.mta.hu, carolyn.doherty@csfk.mta.hu} 
\\[\affilskip]
$^2$Monash Centre for Astrophysics, School of Physics and Astronomy,
Monash University, VIC 3800, Australia \\email:
{\tt amanda.karakas@monash.edu}
\\[\affilskip] 
$^3$Research School of Astronomy and Astrophysics,
Australian National University, Canberra, ACT 2611, Australia
\\[\affilskip] 
$^4$E.A. Milne Centre for Astrophysics, University of Hull, Cottingham Road Hull 
HU6 7RX United Kingdom \\email:
{\tt m.pignatari@hull.ac.uk}
\\[\affilskip] 
$^5$The NuGrid collaboration (www.nugridstars.org)}

\pubyear{2016}
\volume{323}  
\jname{Planetary nebulae: Multi-wavelength probes of stellar and galactic evolution}
\editors{X. Liu, L. Stanghellini \& A. I. Karakas, eds.}

\begin{document}

\maketitle

\begin{abstract}
Planetary nebulae retain the signature of the nucleosynthesis and mixing
events that occurred during the previous AGB phase. Observational
signatures complement observations of AGB and post-AGB stars and their
binary companions. The abundances of the elements
heavier than iron such as Kr and Xe in planetary nebulae can be used to
complement abundances of Sr/Y/Zr and Ba/La/Ce in AGB stars, respectively, to
determine the operation of the {\em slow} neutron-capture process 
(the $s$ process) in AGB stars. 
Additionally, observations of the Rb abundance in Type I planetary nebulae may allow us to 
infer the initial mass of the central star.
Several noble gas components present in meteoritic stardust silicon
carbide (SiC) grains are associated
with implantation into the dust grains in the high-energy environment
connected to the fast winds from the central stars during the planetary
nebulae phase. 

\end{abstract}

\firstsection 
\section{Introduction}

The evolutionary track of the life of a star of 2 \msun\ after core H exhaustion 
proceeds to the red giant phase, until core He burning stars. After core He 
exhaustion, the star becomes a red giant again and climbs the asymptotic giant branch 
(AGB). During the AGB phase, the stars loses most of its envelope via strong 
stellar winds, until mainly the C-O degerate core is left, with a thin layer of H 
($\sim 10^{-3}$ \msun) still burning. At this point the star travels back in the H-R 
diagram towards hotter temperatures. If it becomes hot enough before all the material 
expelled in the previous AGB phase is dispersed, the UV photons from the central star 
illuminate the circumstellar material producing a planetary nebula. It follows that 
in planetary nebulae we observe the chemical abundances produced during the 
AGB phase.
 
During the AGB the structure of the star can be described by two components: the 
entended convective H-rich envelope and the compact core. The core is hot and dense, 
and nuclear reactions can occur there that produce and modify the chemical 
abundances. Two burning shells are present: the H and the He burning shells. These shells 
act alternately, with recurrent episodes of He burning driving convective motions 
(a thermal pulse, TP) in the whole region in-between the two shells (the intershell). 
As the star expands and cools down, H burning is extinguished, the TP ceases, and He 
continues to burn radiatively at the bottom of the intershell. At this point the base of 
the convective envelope may sink deep inside the intershell (third dredge-up, TDU) 
and carry to the stellar surface the products of H and He burning. After the TDU, He 
burning extinguishes and H burning starts up again. This cycle of H burning, TP, and TDU, 
repeats until the envelope mass is lost. The material brought to stellar surface is 
carried into the surroundings by the AGB winds and contributes to making up the chemical 
enrichment of the Galaxy. For reviews on AGB stars, see 
\citet{herwig05} and \citet{karakas14dawes}.

Four main chemical signatures result from the AGB evolution:

\begin{enumerate}

\item{During the TPs, partial He burning makes carbon, and not much oxygen, in the 
intershell and stars in the mass range roughly 1.5 to 4 \msun (depending on the 
initial composition) can eventually become the observed C-rich stars 
\citep[see, e.g., the middle panels of Figs. 2-4 of][]{karakas16}.}

\item{The base of the convective envelope can be hot enough to drive proton captures 
on the isotopes of the elements from C to Al (hot bottom burning, HBB). They transform, 
for example, C into N in stars of initial mass larger than roughly 4 \msun  
\citep[see, e.g., the bottom panels of Figs. 2-4 of][]{karakas16}.}

\item{The ingestion of H-burning ashes in the TPs results in the production of specific 
rare isotopes, such as $^{19}$F and $^{22}$Ne \citep[e.g.,][]{lugaro04,abia15}.}

\item{Finally, AGB stars are the site of origin of half of the cosmic abundances of 
the elements heavier than iron, via the operation of the $slow$ neutron capture 
process (the $s$ process) in the intershell.}

\end{enumerate}

The last of these four points will be discussed in more detail in the present 
paper, together with implications from the observations of planetary nebulae.

\subsection{The $s$ process}

Because of their high number of protons ($>$26) the elements heavier than Fe have a 
large Coulomb barrier and can be produced only by capturing neutrons. The $s$ process 
occurs typically in hydrostatic conditions at relatively low neutron densities 
(typical 10$^{8}$ n/cm$^{3}$, but up to 10$^{13}$ n/cm$^{3}$ in the more massive AGB 
stars). The path of neutron captures proceeds mostly through stable nuclei, since 
unstable nuclei decay to their corresponding isobar before they can capture another 
neutron.

\begin{figure}[h]
\begin{center}
 \includegraphics[width=3.4in]{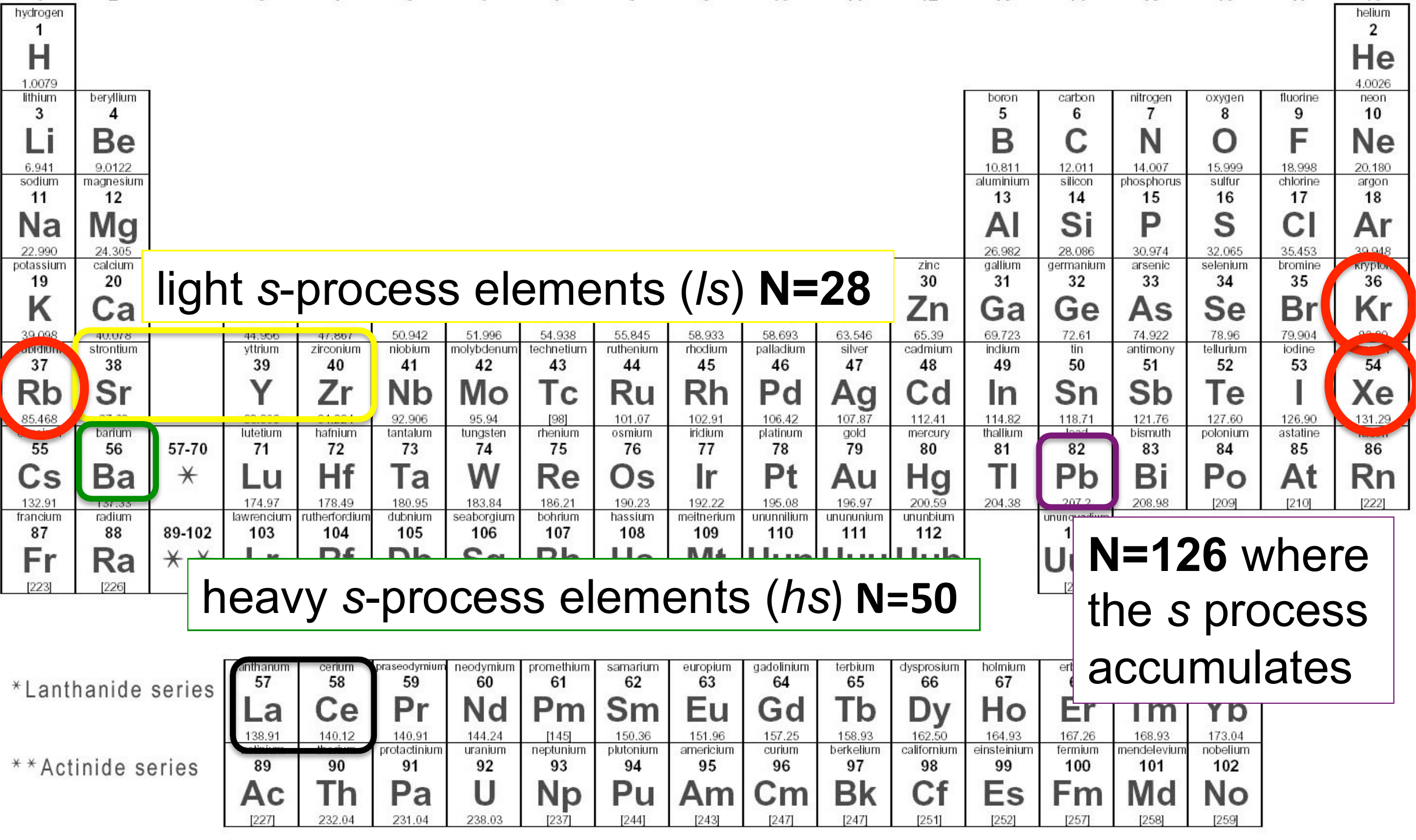} 
 \caption{
Periodic table of the elements with the locations of the first, second and third 
$s$ process peaks marked along with the corresponding neutron numbers.
Also highlighted are the noble gases Kr and Xe and the element Rb, which can be 
observed in planetary nebulae.}
   \label{fig1}
\end{center}
\end{figure}

Along the path of stable nuclei, the flux of neutron captures encounters three {\em 
bottlenecks} corresponding to the magic number of neutrons N = 50, 82, 126 
(Fig.~\ref{fig1}). Nuclei with such numbers of neutrons are relatively more stable against neutron 
captures than other nuclei, and tend to accumulate until the neutron flux  
reaches a large enough value to make the probability of neutron captures on
these nuclei significant. The total, time-integrated neutron flux is the crucial 
quantity that determines if the bottlenecks are bypassed and if the production flux can 
proceed to the next bottleneck. The bottleneck nuclei determine the final $s$-process 
distribution and correspond to the so-called first $s$-process peak -- or {\em light} 
s-process elements ({\em ls}) with N=50, Sr, Y, Zr; second $s$-process peak -- or 
{\em heavy} s-process elements ({\em hs}) with N=82, Ba, La, Ce; and third 
$s$-process peak at N=126 (Pb), which can only accumulate since the $s$-process path 
ends there. The traditional $s$-process {\em indexes} are defined as 
[ls/Fe]=([Sr/Fe]+[Y/Fe]+[Zr/Fe])/3 and [hs/Fe]=([Ba/Fe]+[La/Fe]+[Ce/Fe])/3; although
other choices have also been used in the literature \citep[see discussion in][]{lugaro12}.

\subsection{The $s$-process in AGB stars}

In the intershell of AGB stars free neutron are produced by two reactions. The 
$^{13}$C($\alpha$,n)$^{16}$O reaction is the main neutron surce in AGB stars of mass 
lower than roughly 4 \msun. A region rich in $^{13}$C, the $^{13}$C {\em pocket}, is 
assumed to form via mixing of protons from the envelope into the intershell at the 
deepest extent of the TDU \citep{gallino98,goriely00,lugaro03a,cristallo09}. The 
protons react with the abundant $^{12}$C to produce $^{13}$C via 
$^{12}$C(p,$\gamma$)$^{13}$N($\beta^+$)$^{13}$C. The $^{13}$C pocket controls the 
total number of neutrons and thus the observed $s$-process distribution, i.e., the 
[hs/ls] ratio. The 
absolute $s$-process abundances [ls/Fe] and [hs/Fe] are controlled also by the TDU and other 
dilution effects. For example, binary transfer plays a crucial role 
in interpreting the data derived from an AGB 
star companion (a so-called {\em extrinsic} $s$-process enhanced star), rather than an
AGB star itself (an {\em intrinsic} $s$-process enhanced star). Because both the formation 
of the $^{13}$C pockets and the effects of 
stellar rotation and magnetic fields on the 
neutron flux are very uncertain, observations of [hs/ls] represent one of a few viable 
ways to understand the $s$ process in low-mass AGB stars.

For AGB stars more massive than roughly 4 \msun, the main source of neutrons is 
instead the $^{22}$Ne($\alpha$,n)$^{25}$Mg reaction, which operates at higher 
temperatures (above 300 MK) inside the convective TPs. This neutron source produces a 
lower total number of neutrons but a higher neutron density than the $^{13}$C source. 
As such, it controls the possible operation of branching points, introducing deviations from 
the main $s$-process path. For instance, the abundance ratio [Rb/Zr] is an 
indicator of the neutron density and of the activation of the $^{22}$Ne source. 
The production of Rb 
is much increased when the neutron magic isotope $^{87}$Rb (N=50) is reached by the 
$s$-process path via the activation of the branching 
points at the unstable $^{85}$Kr and $^{86}$Rb \citep{abia01,vanraai12}. Large 
uncertainties are present in the operation of the $^{22}$Ne 
neutron source 
including the rate of the neutron source reaction, the efficiency of the 
TDU, and the mass loss rate in massive AGB stars.
The difficulty in constraining the operation of this neutron source using
observations is that the [Rb/Zr] ratio in massive AGB stars is highly dependent on 
the model atmosphere used to analyse the spectra \citep{zamora14}. Independent 
constraints are crucial to settle the question of how much Rb is 
produced by massive AGB stars.

\section{New directions}

Observations of the elements heavier than Fe in planetary nebulae (see Sterling et al., 
these proceedings) provide us complementary indicators and independent constraints for 
the $s$-process in AGB stars. As it can be seen from Fig.~\ref{fig1} the noble gases 
Kr and Xe, which are observable in planetary nebulae, are located very close to the 
first and second $s$-process peaks, respectively. Interestingly, only a small fraction of the 
cosmic abundances of these noble gases are produced by the $s$-process in AGB stars
\citep{arlandini99,bisterzo14}: Xe is a 
typical element made by the $rapid$ neutron captures in explosive environments 
\citep{thielemann11}. Krypton, on the other hand, is 
mostly produced 
in massive stars by a multitude of processes, including the weak $s$ process 
and different explosive mechanisms \citep[e.g.,][and references therein]{roberts10,pignatari16}.

Nevertheless, the $s$ process in AGB stars contributes 
roughly 30\% and 15\% of the solar Kr and Xe, respectively \citep[][]{bisterzo14}. 
This production, although much lower in 
absolute terms, follows the production of the $s$-process first and second peaks. This 
means that effectively the $s$-process models predict [Xe/Kr] values comparable to [hs/ls] 
values. The observed [Xe/Kr] ratios in planetary nebulae can be used to 
further constrain the total number of free neutrons and consequently the \iso{13}C 
pocket \citep[][]{sharpee07,pignatari08}. 
Furthemore, Rb can be directly seen in planetary nebulae, providing us with a 
constrain on the initial mass of the parent star and the opportunity to disentangle the 
effects of hot bottom burning in massive AGB stars (potentially Rb-rich) and of {\it 
extra} mixing in low-mass AGB stars (certainly Rb-poor) on the He and N excesses observed 
in Type I planetary nebulae.

\begin{figure}[h]
\begin{center}
 \includegraphics[width=3.4in]{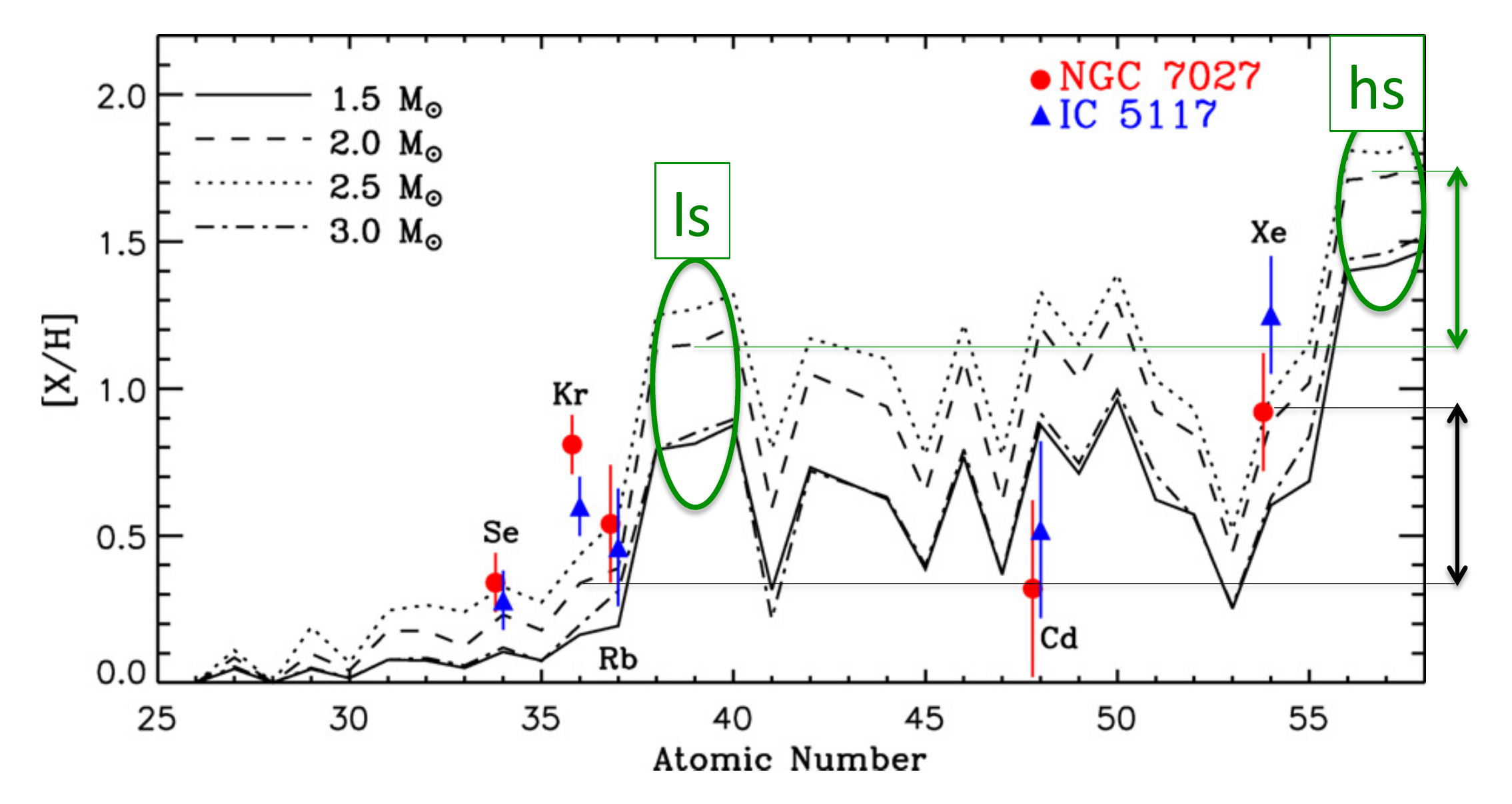} 
 \caption{Figure adapted from \citet{sterling16} (\copyright AAS. Reproduced with permission)
showing the comparison between 
abundances observed in two planetary nebulae (red and blue circle symbols, 
with error bars) 
and results from models of the $s$ process in AGB stars from the FRUITY database 
\citep{cristallo15} for different masses and metallicity [Fe/H]=$-$0.37. 
Also indicated are the 
location of the first and second $s$-process peaks, labelled as ls and hs, respectively. 
The green and black arrows on the right side of the figure measure the [hs/ls] and the
[Xe/Kr] ratios, respectively, predicted by the 2 \msun\ model. It can be seen that 
[hs/ls]$\simeq$[Xe/Kr].}
   \label{fig2}
\end{center}
\end{figure}

From a preliminary analysis by \cite{pignatari08}, the observations of the 
s-process [Xe/Kr] in PNe seem 
to be consistent with the theoretical predictions of AGB models. 
However, \cite{sterling16} (see Fig.~\ref{fig2}) show a mismatch between 
models and observations for Kr, with the observations being higher than the model 
predictions. This problem needs to be analyzed in more details, also in the light 
of the other independent s-process observations of noble gases available
from meteoritic presolar silicon carbide 
(SiC) grains, which formed in the winds of C-rich AGB stars.

\subsection{The record of planetary nebula processes in stardust SiC}

\begin{figure*}
   \begin{center}
 \includegraphics[height=6.5cm]{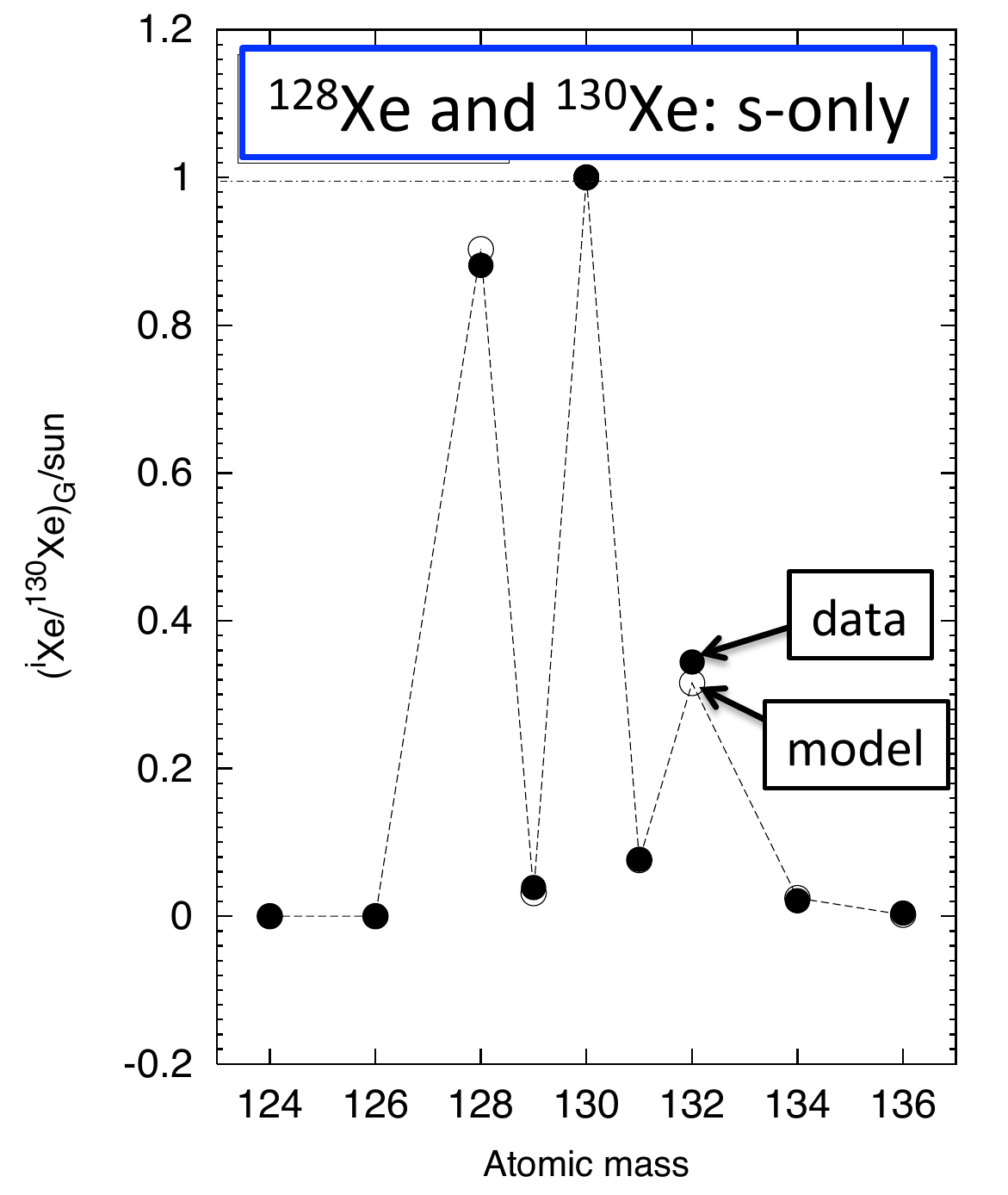}
 \includegraphics[height=6.5cm]{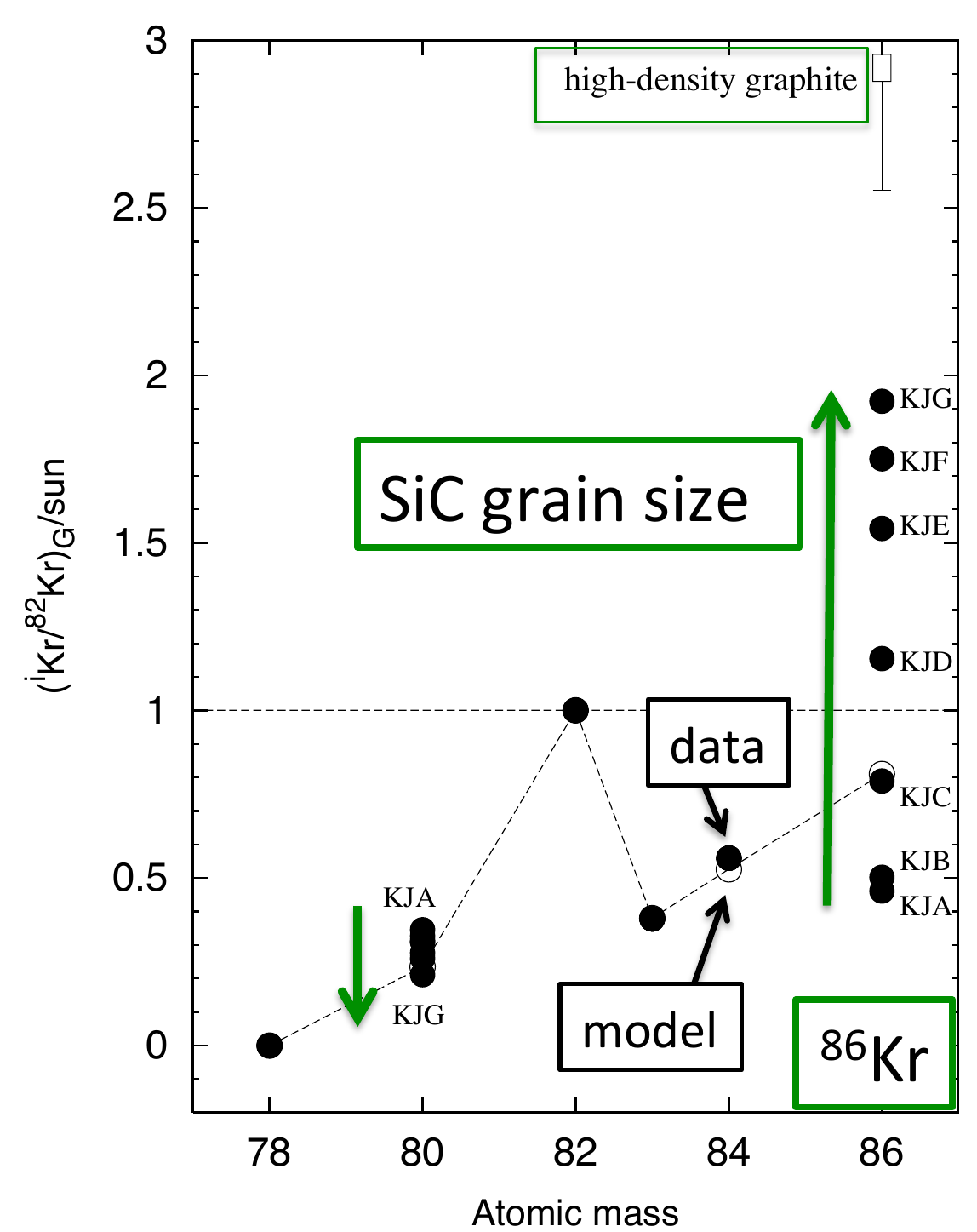}\\
   \end{center}
\caption{Figures adapted from \citet{lugaro05} showing a
comparision between the isotopic composition of Xe and Kr (normalised to 
\iso{130}Xe and \iso{82}Kr, respectively, 
and to the solar composition) measured in collections of millions of 
SiC grains \citep[in {\em bulk} by][]{lewis94} and 
predicted by AGB models for the typical parent stars 
of the grains (1.5--3 \msun\ and solar metallicity).} 
\label{fig3}
\end{figure*}

Meteoritic stardust SiC grains contain noble gases that carry the signature of processes 
occurring in AGB stars and planetary nebulae. One of the first evidences that 
most of these grains originated from the envelopes of C-rich AGB star comes from the 
composition of the noble gas Xe (left panel of Fig.~\ref{fig3}), which shows the indisputable 
signature of the $s$ process in AGB stars. The Xe isotopic ratios produced 
by the $s$ process are predominantly controlled by the neutron-capture 
cross sections of the different Xe isotopes, which strongly favour the production
of \iso{128}Xe and \iso{130}Xe. These are the two Xe $s$-only isotopes that are 
shielded from $r$-process production by their isobars in Te and are predominantly 
produced by the $s$ process.

\begin{figure}[h]
\begin{center}
 \includegraphics[width=3.4in]{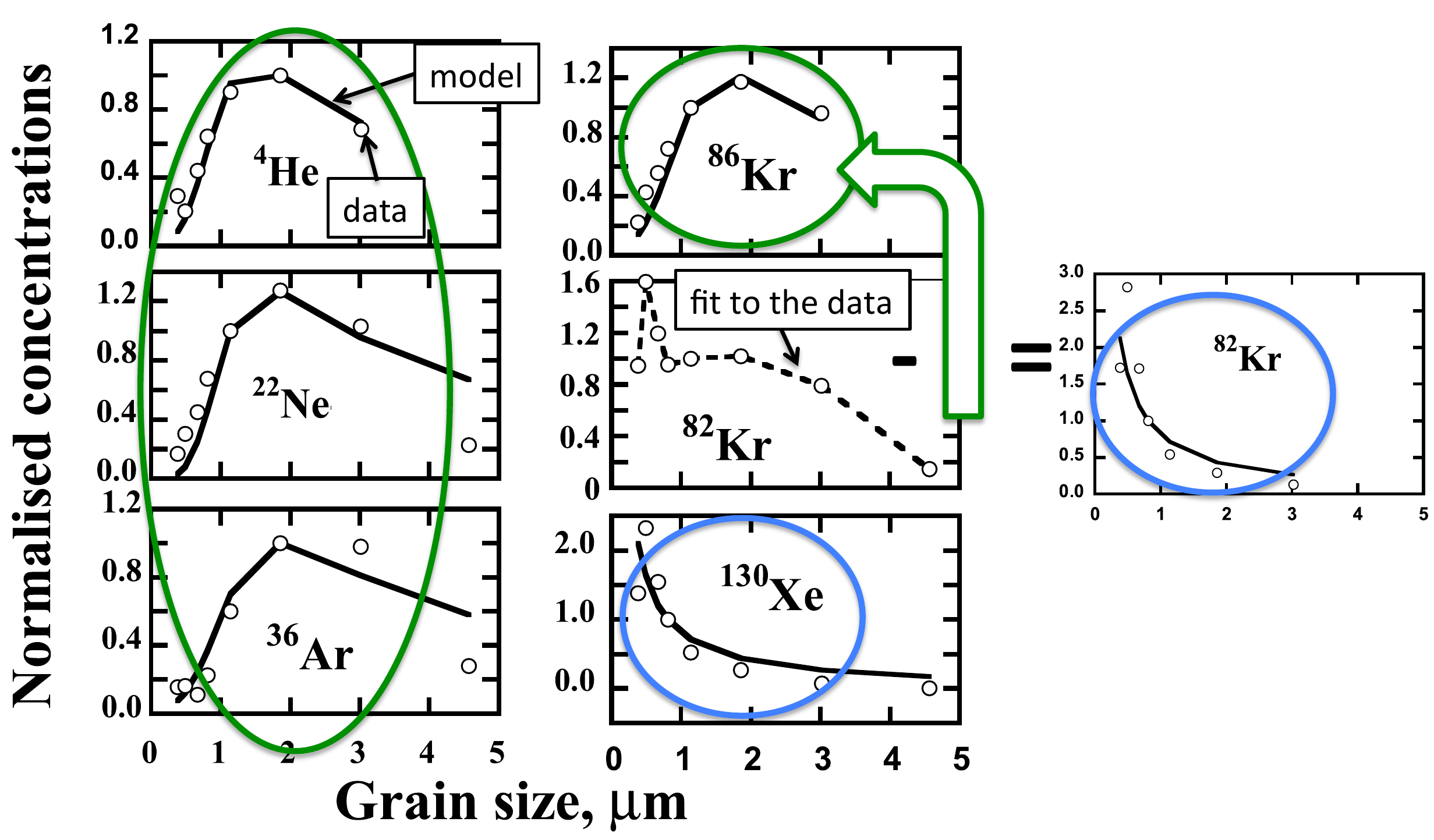} 
 \caption{Figure adapted from \citet{verchovsky04} 
(\copyright AAS. Reproduced with permission)
and showing a comparison between 
the abundances of the different noble gases in meteoritic SiC as a function of the grain 
size \citep[from][]{lewis94} and the implantation model. Two clearly different patterns 
can be identified: \iso{4}He, \iso{22}Ne, \iso{36}Ar, and \iso{86}Kr are most abundant in 
large grains (highlighted in green), \iso{130}Xe is most abundant in the small grains
(highlighted in blue), while the \iso{82}Kr 
abundance is made by a mixture of the two patterns, i.e., the blue pattern 
can be obtained
by subtracting the \iso{86}Kr green pattern from the total \iso{82}Kr.}
   \label{fig4}
\end{center}
\end{figure}

Very intriguing is the abundance trend of the different noble gases, He, Ne, 
Ar, Kr, and Xe with the size of the grains. Since noble gases are chemically 
inert, they are believed to have been incorporated in the SiC grains via an 
implantation process. 
\citet{verchovsky04} analysed the data from \cite{lewis94} in 
the light of an implantation model and discovered that the two 
different patterns (highlighted in green and blue in Fig.~\ref{fig4}) can be reproduced by 
the models by varying the energy per nucleon, an input parameter of the model. 
A relatively low energy per nucleon, of the order of 0.5 keV, can reproduce the pattern 
observed in Xe. This energy can be associated with turbulence in the winds during 
the AGB phase, which have typical speeds of 10-30 km/s (the {\em cold} component, 
Table~\ref{tab1}). As 
a consequence, Xe atoms must have been implanted carrying the signature of the matter 
in the winds of the AGB star. On the other hand, a much higher energy per nucleon 
($\simeq$40 keV) is required to reproduce the pattern observed in He, Ne, Ar. This is 
because if the atom has too much energy, it cannot be stopped inside small grains. 
Post-AGB/planetary nebula fast winds ($\simeq$700 km/s) 
can imprint such an energy in the noble gases atoms (the {\em hot} component, 
Table~\ref{tab1}). As a 
consequence, He, Ne, and Ar must have been implanted carrying the signature of matter 
in the post-AGB winds, the same winds invoked as one of the 
possible factors involved in the production of planetary nebulea.
Finally, Kr is a mixute of the two components, some of it being implanted in 
the AGB phase, and some during the planetary nebula phase.

\begin{table}
  \begin{center}
  \caption{Qualitive description of the two components of noble gases 
implanted in meteoritic SiC grains.}
  \label{tab1}
 {\scriptsize
  \begin{tabular}{|c|c|c|}
\hline 
{\bf Component} & {\bf cold} & {\bf hot} \\
\hline 
Grain size ($\mu$m) & $<1$ & $>1$ \\
\iso{4}He & no$^1$ & yes \\
\iso{22}Ne & no$^1$ & yes \\
\iso{36}Ar & no$^1$ & yes \\
\iso{82}Kr & yes & yes \\
\iso{86}Kr & no$^2$ & yes \\
\iso{130}Xe & yes & no$^3$ \\
Energy/nucleon (keV) & 0.5 & 40 \\ 
Implantation winds & AGB & post-AGB \\
Wind speed (km/s) & 10-30 & 700 \\
AGB nucleosynthetic signature & envelope & intershell (final) \\ 
\hline 
  \end{tabular}
  }
 \end{center}
\vspace{1mm}
 \scriptsize{
 {\it Notes:}\\
  $^1$Not ionised.\\ 
  $^2$Nucleosynthetic signature not present.\\ 
  $^3$Much less abundant than the cold component.}
\end{table}

The question 
arises of why He, Ne and Ar show only the fast wind/planetary 
nebula component, Xe shows only the AGB wind component, and Kr shows them both. 
\citet{verchovsky04} 
proposed that these signatures are due to the
different ionisation energy of the noble gases, which decreases with the number of 
electrons: Xe has the lowest ($\simeq$ 12 eV) ionization energy, and He the highest 
($\simeq$ 25 eV) ionisation energy. If we 
assume that the atoms must be ionized to be implanted in the 
grains, then their different ionization energies can fractionate their abundances. 


From the observed isotopic ratios of the noble gases we can gather detailed information 
about the nucleosynthetic signature carried by the two components. The Xe isotopic 
ratios are matched by the composition predicted in the AGB winds (left panel of 
Fig.~\ref{fig3}). 
This is the mixture of $s$-process-rich intershell material carried to the envelope by the 
TDU during the whole AGB phase and the initial composition of the stars, close to the 
solar system composition. On the other hand, to match the nucleosynthetic signatures 
presented by He and Ne, and in particular the so-called Ne-E(H) component extremely 
enriched in \iso{22}Ne, almost pure intershell material is required, with no dilution by the
original envelope material. Post-AGB winds can carry such almost pure 
signature only in the case some flavour of a late or very late TP occurred 
\citep{bloecker01}, which affects $\sim$20\% of the AGB population \citep[e.g.,][]{werner06}. 
It is 
possible that a nucleosynthetic signature such as that of the H-rich envelope 
is also present for He and Ne in the hot component, but, it is not seen because its 
abundance is much lower than that of the intershell in the late or very late TP scenario.

The \iso{82}Kr abundance shows both the cold and the hot components (as
Kr may be ionised in both cases), while \iso{86}Kr shows only the fast wind/planetary 
nebula hot component. 
The most likely explanation is that the nucleosynthetic signature of the cold component 
carries a \iso{82}Kr but not a \iso{86}Kr excess, while the nucleosynthetic signature of 
the hot component carries both. The explanation for this difference lies in the fact that \iso{82}Kr is located on the main $s$-process path, 
while \iso{86}Kr is produced by the activation of the branching point at \iso{85}Kr, for 
neutron densities exceeding few 
$10^8$ n/cm$^3$. 
The hot component recorded in large grains points to high 
\iso{86}Kr/\iso{82}Kr ratios in the intershell (right panel of Fig.~\ref{fig3}). 
The stellar origin of such 
signature remains debated. Potential proposed explanations are models of metallicity lower 
than solar and either a higher efficiency of the activation of \iso{13}C pocket 
\citep{pignatari06} or the ingestion of the \iso{13}C pocket inside the TPs \citep{raut13}, 
which can occur if the initial mass is lower than 1.5 \msun. 
However, such models predict Sr and Ba isotopic ratios inconsistent with
measurements of the same bulk and single SiC grains 
\citep{gallino97,lugaro03b,barzyk07,avila13,liu15}.
The impact of convective-boundary mixing processes at the bottom of 
the convective thermal pulse \citep[e.g.,][]{lugaro03a,battino16} needs to be also 
evaluated for Kr and Xe.  
Another possibility would be that the production 
of extra \iso{86}Kr occurs during the post-AGB phase in 
association with a late or a very late TP
that is already required to carry to the 
stellar surface the almost pure intershell material seen in the hot component
\citep[e.g., as in the case of the
{\em intermediate} neutron-capture process seen in Sakurai's object;][]{herwig11}.

Finally, it would be important to fully define the connection between the excess \iso{86}Kr in the 
stardust data and the Kr excess seen in the planetary nebulae observations 
(Fig.~\ref{fig2}). From a back-of-the-envelope calculation, if we increase by a factor 
of 4 the \iso{86}Kr abundance in the 2.5 \msun\ model shown in Fig.~\ref{fig2}, to 
simulate an extreme 
composition such as that shown by the high-density graphite grains (Fig.~\ref{fig3}), 
the [Kr/Fe] ratio
would increase from its current value of 0.43 to 0.66, 
which is a much better match with the 
data. 
As we mentioned ealier, the capability of theoretical AGB models to reproduce 
PN $s$-process observations of Kr and Xe is still matter of debate, and a consistent scenario should be derived that works for both PNe and measurements from presolar grains. 

\section{Summary and conclusions} 

Nucleosynthesis in AGB stars shapes the abundances observed in planetary nebulae via 
a complex interplay of nuclear reactions and mixing processes 
such as the third dredge-up and hot bottom burning. 
As such, planetary nebula abundances provide 
an independent, complementary testbed for all these processes, including the production of the elements heavier than rion via the $s$ process. 
Conversely, the observed chemical abundance can be used to constraint 
the evolutionary path leading to the formation of planetary nebulae and the 
initial mass range.   
Specifically, we confirm that observations of [Xe/Kr] 
are a potential proxy for the traditional [hs/ls] $s$-process index and that the abundance 
of Rb can provide us crucial further constraints on the initial masses of planetary 
nebulae and the nucleosynthesis and mixing processes in massive AGB stars. Furthermore, 
noble 
gas abundances observed in meteoritic stardust grains are tracers of processes in 
planetary nebulae, from ionisation and implantation, to nucleosynthesis in the central 
star. 
The combination of all these different independent constraints will allow us to 
open new ways to understand planetary nebulae and the link to their parent AGB stars.

\bibliographystyle{apj}
\bibliography{apj-jour,library}


\end{document}